\documentclass{article}
\usepackage{graphicx} 
\usepackage{amsmath}
\usepackage{amssymb}
\usepackage{cite}
\usepackage{float}
\usepackage{hyperref}
\usepackage{geometry}
\title{Fractional Holographic Dark Energy}
\author{%
    Oem Trivedi $^{1}$\thanks{oem.t@ahduni.edu.in, \href{https://orcid.org/0000-0002-2086-4127}{ORCID:0000-0002-2086-4127}}, Ayush Bidlan $^{2}$\thanks{i21ph018@phy.svnit.ac.in , \href{https://orcid.org/0000-0002-2842-3334}{ORCID:0000-0002-2842-3334}}, Paulo Moniz $^{3}$ \thanks{pmoniz@ubi.pt, \href{https://orcid.org/0000-0001-7170-8952}{ORCID:0000-0001-7170-8952} }
}

\date{%
    \small
    $^{1}$International Centre for Space and Cosmology, Ahmedabad University, Ahmedabad 380009, India\\
     $^{2}$Department of Physics, Sardar Vallabhbhai National Institute of Technology, Surat, Gujarat 395007, India\\
    $^{3}$Departamento de Física, Centro de Matemática e Aplicações (CMA-UBI), Universidade da Beira Interior, Rua Marquês d’Avila e Bolama, 6200-001 Covilhã, Portugal\\
    \today 
}

\begin{document}

\maketitle

\begin{abstract}
    Holographic dark energy theories present a
    fascinating interface 
    to probe
    late-time cosmology, as guided by contemporary ideas about quantum gravity.
    In this work, we present a new holographic dark energy scenario 
    designated
    ``Fractional Holographic Dark Energy"(FHDE). This model extends the conventional framework of HDEs by incorporating specific features 
    from fractional calculus recently applied, e.g., in 
    cosmological settings. In this manner, we 
    retrieve 
    a novel form of HDE energy density. We then show how FHDE can provide a consistent picture of the evolution of the late-time universe even with the simple choice of the Hubble horizon as the IR (infrared) cutoff. We provide detailed descriptions of the cosmological evolution, showing how the fractional calculus ingredients 
    can alleviate quite a few issues associated with the conventional HDE scenario. Concretely, we compute and plot diagrams using the Hubble horizon cutoff for HDE. The density parameters for DE and dark matter (DM), the deceleration parameter, and the DE EoS parameter indicate how the universe may evolve within our ``FHDE" model, fitting within an appropriate scenario of late-time cosmology.
\end{abstract}

\section{Introduction}
The discovery of the universe's late-time acceleration is a defining moment in cosmology \cite{SupernovaSearchTeam:1998fmf}. This unexpected revelation unveiled a significant discrepancy between our understanding of gravity and the observed expansion history of the cosmos. Since then, the pursuit of understanding this phenomenon has fueled a vibrant field of research encompassing a diverse array of theoretical frameworks. The conventional approach to address late-time cosmic acceleration centres on introducing a Cosmological Constant within the framework of General Relativity \cite{Weinberg:1988cp,Padmanabhan:2002ji}. This term acts as a form of ``dark energy" (DE), a mysterious entity contributing a repulsive force counteracting attractive gravity and driving accelerated expansion. However, the $\Lambda$CDM model \cite{Perivolaropoulos:2021jda,Condon:2018eqx} faces challenges in explaining the origin and fine-tuning required for the specific value of the cosmological constant.
\\
\\
Alternative explanations for late-time acceleration venture beyond the realm of General Relativity. Modified gravity theories propose alterations to the gravitational field equations themselves, potentially leading to an explanation for the observed expansion without resorting to dark energy \cite{Nojiri:2010wj,Nojiri:2017ncd}. Additionally, scalar field models posit the existence of a new dynamical field, quintessence or phantom energy, whose dynamics may drive the late-time acceleration \cite{Zlatev:1998tr,Faraoni:2000wk,Capozziello:2002rd,Odintsov:2023weg}. The quest for a more fundamental understanding of gravity has also led to the exploration of late-time acceleration in the context of quantum gravity theories. String theory frameworks like Braneworld cosmology, loop quantum cosmology, and asymptotically safe cosmology offer intriguing possibilities by incorporating quantum mechanical effects into the description of spacetime \cite{Sahni:2002dx,Sami:2004xk,Tretyakov:2005en,Chen:2008ca,Fu:2008gh,Bonanno:2001hi,Fernandes:2011dd,Andrianov:2008az,Gorini:2002kf,Kamenshchik:2012ij,Keresztes:2009vc}.
Despite the plethora of proposed solutions, fundamental questions regarding the nature of dark energy and late-time acceleration persist. Moreover, a critical current challenge lies in reconciling conflicting measurements of the Hubble constant, known as the Hubble tension \cite{Planck:2018vyg,riess2019large,riess2021comprehensive}. This discrepancy highlights potential shortcomings in our current understanding of the Universe's expansion history and the underlying physics governing it. It underscores the need for advancements in gravitational physics to bridge the gap between theoretical frameworks and observational data. The recent findings from the Dark Energy Spectroscopic Instrument (DESI) \cite{adame2024desi3,adame2024desi4,adame2024desi6} further emphasize the need for a more nuanced understanding of dark energy. These observations hint at potential deviations from the $\Lambda$CDM model, suggesting that alternative explanations for late-time cosmic acceleration deserve continued exploration.
\\
\\
In this context, applying the holographic principle to cosmology emerges as a promising avenue for exploration \cite{tHooft:1993dmi,Susskind:1994vu}. This principle, arising from string theory and quantum gravity considerations, proposes 
an intriguing
relationship between a system's entropy and its boundary surface area rather than its volume \cite{Bousso:1999xy}. This 
captivating concept offers a potential link between the enigmatic nature of dark energy and the underlying structure of spacetime at the quantum level \cite{Cohen:1998zx,cohen1999effective,wang2017holographic}. Building upon the pioneering work in \cite{Cohen:1998zx}, our
research delves deeper into the theoretical framework of holographic dark energy (HDE).  
In particular, we explore the connection between short-distance cutoffs arising from quantum field theory and long-distance cutoffs imposed by black hole formation constraints. Specifically, in the work of Cohen et al  \cite{cohen1999effective}, the authors showed that holographic considerations led to the inequality \begin{equation} \label{holorelation}
    L^3 \Lambda^3 \leq S^{3/4},
\end{equation}
where S is the system entropy while L and $\Lambda$ denote the IR and UV cutoffs, respectively, and we are working in units such that we take the Planck mass as $m_{p}=1$. Using the Bekenstein-Hawking definition for the entropy, $S \propto A$, 
together with 
the idea that the vacuum energy density $\Lambda^4$ corresponds to the energy density for dark energy, 
i.e., 
$\rho_{DE} \sim \Lambda^4$, we can then arrive at the following expression
\begin{equation}\label{simphde}
\rho=3 c^2 L^{-2}.
\end{equation} 
Quite a few 
other works have explored holographic dark energy from various similar aspects in recent years \cite{Nojiri:2017opc,Oliveros:2022biu,Granda:2008dk,Khurshudyan:2016gmb,Wang:2016och,
Khurshudyan:2016uql,Belkacemi:2011zk,
Zhang:2011zze,Setare:2010zy,Nozari:2009zk,Sheykhi:2009dz,
Xu:2009xi,Wei:2009au,Setare:2008hm,Saridakis:2007wx,Setare:2006yj,
Felegary:2016znh,Dheepika:2021fqv,Nojiri:2005pu,Nojiri:2021iko,Nojiri:2020wmh}. In recent years, 
various other alternative forms of HDE have been proposed. 
For example, Tsallis HDE models are based on Tsallis' corrections to the standard Boltzmann-Gibbs entropy (obtained from applying 
these 
corrections
to  black hole physics), 
resulting in
the equation 
\begin{equation} \label{rtsa}
    \rho_{\Lambda} = 3 c^2 L^{-(4 - 2\sigma)},
\end{equation} where $\sigma$ is the Tsallis parameter, which is considered to be positive \cite{Tavayef:2018xwx} and we recover simple HDE in the limit $\sigma \to 1$.  On the other hand, Barrow's modification of the Bekenstein-Hawking formula led to the creation of Barrow HDE models and such models are described by the energy density \begin{equation} \label{rbar}
    \rho_{\Lambda} = 3 c^2 L^{\Delta - 2},
\end{equation} where $\Delta$ is the deformation parameter \cite{Saridakis:2020zol}, which can have a maximum value of $\Delta = 1 $ and in the limit of $\Delta \to 0$ one recovers the simple HDE. There are many other models like the Renyi and Kaniadakis models \cite{Drepanou:2021jiv,moradpour2018thermodynamic} which have been forward in recent years based on various motivations, ranging from thermodynamics to generalized uncertainty principles amongst other approaches \cite{barrow1992cosmology,maggiore1993generalized,tawfik2014generalized}. 
In our work, we would like to propose a new model of dark energy which is based on considerations of fractional calculus. 
(and recent applications in gravitational and (quantum) cosmological scenarios). 
In Section II we 
motivate this scenario and provide the new energy density for the HDE 
whereas  in Section III, we 
investigate in detail the cosmological evolution of our case study. We 
conclude and discuss our work in Section IV.

\section{Fractional Holographic Dark Energy}

Given our interest in the implications of fractional 
calculus
in either 
classical or quantum 
cosmological scenarios 
it is appropriate to provide a brief overview of 
their features.
Fractional calculus extends differentiation and integration to any real or complex orders applicable to the study of 
complicated
dynamical systems and various intricate physical phenomena. Concretely, the fractional derivative generalizes the 
usual integer 
order of differentiation to rational,  real or complex numbers, with several types having been defined, such as the the Liouville, Riemann, Caputo, and Riesz fractional derivatives \cite{herrmann2011fractional}. While no single such derivative type can fully be applied to every situation, their suitability with reality 
depends on the specific problem at hand. The interested reader can refer to various interesting works by Ortigueira for more details \cite{ortigueira2011fractional,ortigueira2017fractional,ortigueira2012relation,ortigueira2023fractional,valerio2023variable,bengochea2023operational,ortigueira2024factory}.
\\
\\
Within fractional calculus, a particularly  enticing  
application is quantum mechanics. Let us remind 
Brownian motion
where the trajectories are self-similar, non-differentiable lines 
with fractal dimensions differing from their topological dimensions. In this context, Feynman and Hibbs re-established 
quantum mechanics as a path integral over Brownian paths using the concept of fractality \cite{feynman2010quantum}. 
In recent years, this led to the development of a fractional path integral, establishing space-fractional quantum mechanics (FQM) as a path integral over another type of steps, within Lévy flight paths, characterized by the Lévy index $\alpha$.
The Gaussian process or Brownian motion is recovered in a specific limit where $\alpha = 2$. A significant result of  FQM is the space-fractional Schrödinger equation (SE), where the second-order spatial derivative of the ordinary SE is replaced by a fractional-order derivative, specifically the quantum Riesz fractional derivative. For further reading, the reader is directed to the monograph and influential works of Nick Laskin on FQM \cite{laskin2000fractional,laskin2002fractional,laskin2000fractional1,laskin2000fractals,laskin2017time,wang2007generalized}.  Additionally, Naber established the time-fractional SE \cite{naber2004time}, replacing the first-order time derivative of the ordinary SE with a fractional-order derivative, concretely the Caputo fractional derivative, while keeping the spatial derivative unchanged. This subsequently led to the development of the spacetime-fractional SE, wherein the corresponding fractional derivatives replace both the second-order spatial and first-order time derivatives. For further readings on the history of fractional calculus, Levy paths and associated interests, one can have a look at \cite{ross2006brief,herrmann2011fractional,samko1993aa,kyprianou2007introduction,mandelbrot1960pareto,levy1925calcul,moniz2020fractional}. 
\\
\\
Regarding its application to cosmology and gravity, fractional calculus can extend any gravitational or cosmological framework to another with non-integer orders in the derivatives of either the classical or quantum regime. Among the extended cosmological frameworks, fractional cosmology is particularly promising for addressing open problems such as the Hubble tension, synchronization, and cosmological constant problems. This approach offers an effective means of investigating gravitational and cosmological phenomena like the evolution of the universe, the dynamics of black holes, and gravitational waves \cite{roberts2009fractional,jamil2012fractional,torres2020quantum,gonzalez2023exact,leon2023cosmology,Marroquin:2024ddg,Micolta-Riascos:2023mqo,Landim:2021ial,Landim:2021www,Socorro:2023xmx,Fumeron:2023sqz,shchigolev2011cosmological,Junior:2023fwb,Rasouli:2022bug}.
\\
\\
Classical cosmological models based on fractional derivatives have typically followed two main approaches: Last-step modification, where fractional ones replace ordinary derivatives in the field equations of the model after obtaining the field equations; and First-step modification, where a fractional derivative geometry is first established, followed by the proposal of a fractional-order action. The first-step modification is considered more fundamental.
The fractional quantum cosmological framework 
has mainly explored how one can derive the fractional Wheeler–DeWitt (WDW) equation, inspired by the space-fractional Schrödinger equation, using the quantum Riesz fractional derivative for a quantum gravity model. A direction where fractional cosmology was found of considerable interest is elaborated in the next paragraph.
 \\
 \\
 It is widely known that entropy can be modified in the context of various theories. 
We have seen examples of 
entropies advanced within
the Barrow and Tsallis HDEs, previously discussed 
in
\cite{Barrow:2020tzx,tsallis2013black}. There has also been quite some work on 
entropies and their cosmological implications besides 
these.
Being more concrete, this has been motivated by various approaches ranging from loop quantum gravity and string theory to generalised uncertainty principles and rainbow gravity \cite{sheykhi2010thermodynamics,cai2008corrected,awad2014minimal,salah2017non,sefiedgar2017entropic,feng2018rainbow}.\\
\\
Recently, Jalalzadeh et al. \cite{jalalzadeh2021prospecting} investigated the effects of fractional quantum mechanics (FQM) on Schwazrschild black hole thermodynamics. The authors 
used a space-fractional derivative of second order (Riesz derivative) and obtained fractional black hole entropy from a modified Wheeler-DeWitt equation. To obtain the modified Wheeler-DeWitt equation, they first used the canonical quantization procedure for a Schawarzschild black hole, by which they ended up at the corresponding Hamiltonian. After this, they included the quantum Riesz derivative in the momentum operator of the Hamiltonian, which led them to the fractional Wheeler-DeWitt equation. The fractional entropy was then found out as \begin{equation} \label{fracentropy}
     S_h = C A^{\frac{2+\alpha}{2\alpha}} ,
 \end{equation} where \(\alpha\) is a fractional parameter therein  constrained 
 to $1 < \alpha \leq 2$ and the standard case is recovered for \(\alpha = 2\);  
 $C$ is some constant which can produce an appropriate combination of Planck units to reproduce the standard Bekenstein-Hawking relation in the $\alpha=2$ case. This equation implies that the entropy is a power-law function of its area and  
resembles the Barrow and Tsallis entropies, although they have different motivations and physical principles. Using the holographic inequality \eqref{holorelation} and considering the entropy \eqref{fracentropy}, one ends up with \begin{equation}
    \Lambda^3 L^3 \leq (C A^{\frac{2+\alpha}{2\alpha}})^{\frac{3}{4}}.
\end{equation}
From the consideration that $\rho_{DE} \sim \Lambda^4$, one can consider an  energy density for Holographic dark energy, based on fractional calculus features and associated quantum effects,  
as \begin{equation}
    \rho_{DE} = \gamma L^{\frac{2 - 3 \alpha}{\alpha}},
\end{equation}
where $\gamma$ is some constant. One immediately notes that we can recover the standard HDE \eqref{simphde} in the case $\alpha = 2$, which is indeed the limit in which the entropy \eqref{fracentropy} decays to the standard Bekenstein-Hawking case. To be more in line with the original HDE proposal, we can set $\gamma = 3 c^2$, where $c$ is again a constant of $\mathcal{O}(1)$ and after this consideration, we can write 
our proposed  HDE energy density as 
\begin{equation} \label{fracrho}
    \rho_{DE} = 3 c^2 L^{\frac{2 - 3 \alpha}{\alpha}}
\end{equation} 
This forms the basis of our new HDE hypothesis 
and we will now investigate
if this model can provide a consistent scenario of universal evolution in the late-times. One would note that there is a similarity of this formula with the Barrow and Tsallis cases \eqref{rtsa}, \eqref{rbar}, but that is because the form of the entropies in those cases are quite similar and are all of the forms of power laws entropy-area relations. However, as we have made clear in this section, the motivation for this HDE model is quite different than the one for the Tsallis case (namely, arising from modifications in fundamental thermodynamics) and Barrow case (namely, from studies of black hole deformations).  

\section{Cosmological evolution}

We start our analysis by indicating
we will be dealing with a flat FLRW universe, where we shall consider that the contributions of dark energy and dark matter to the universal energy density are dominant. Indeed, this is a reasonable approximation often used in cosmology literature. It 
means that our Friedmann equation takes the form \begin{equation} \label{fried}
    H^2 = \frac{\rho}{3} = \frac{\rho_{DE} + \rho_{m}}{3}.
\end{equation}
The continuity equations for dark matter and dark energy have the usual form as well \begin{equation} \label{contm}
    \Dot{\rho_{m}}+  3 H \rho_{m} (1+ w_{m}) = 0,
\end{equation}
\begin{equation} \label{contd}
    \Dot{\rho_{DE}}+ 3 H \rho_{DE} (1+ w_{DE}) = 0,
\end{equation}
where $w_{m}$ and $w_{DE}$ refer to the equation of state parameters (EoS) for dark matter and dark energy. Note that 
we are not considering an interacting dark sector here. Furthermore, we shall consider that dark matter has a pressureless form, meaning that $w_{m} \sim 0$. Before proceeding further, we must choose the cutoff scale $L$. The early suggestion was considering a cutoff scale given by $L \to H^{-1}$, termed the Hubble horizon cutoff. This choice aimed to alleviate the fine-tuning problem by introducing a natural length scale associated with the inverse of the Hubble parameter $H$, but it was found that this particular scale resulted in the dark energy EoS parameter approaching zero, and it also failed to contribute significantly to the current accelerated expansion of the universe. An alternative idea was then to consider 
the utilization of the particle horizon as the length scale \begin{equation} \label{lp}
    L_{p} = a \int_{0}^{t} \frac{dt}{a}. \quad
\end{equation}  This alternative resulted in an equation of state parameter higher than $-1/3$, but despite this modification, the challenges of explaining the present acceleration remained unresolved.
Another option considered was the future event horizon as the length scale \begin{equation} \label{lf}
     L_{f} = a \int_{a}^{\infty} \frac{dt}{a}.
\end{equation}  Although the desired acceleration regime can be achieved in this case, concerns are raised regarding causality. Another option for the cutoff scale was the Granda-Oliveros cutoff, which took into account the derivative of $H$ into their definition of $L$ as well \cite{Granda:2008dk} \begin{equation}
    L = (\alpha H^2 + \beta \dot{H} )^{- \frac{1}{2}}.
\end{equation} The most generalized cutoff choice is the Nojiri-Odintsov cutoff \cite{Nojiri:2005pu,Nojiri:2017opc,nojiri2019holographic}, where we could have the form \begin{equation} \label{nocutoff}
    L =  L(H,\dot{H},\ddot{H},...L_{p},L_{f}, \dot{L_{p}},\dot{L_{f}}...). 
\end{equation} An additional issue with all these cutoffs relates to the classical stability of these models against perturbations \cite{Myung:2007pn}, an issue seen in all classes of HDE energy densities with various cutoffs. In general, one can have HDE models where the density could be of the form \eqref{nocutoff} being general functions of $H$, particle horizon and event horizon scales. Let us 
note that regarding all the forms of HDEs that we have discussed, whether it is Tsallis, Barrow, others or our own herewith, it could just be said to be such that it would 
approach the form in \eqref{nocutoff}. Although such a comment may be 
mathematically of interest,
one in general 
endeavours 
to construct, in a justified, proper manner, 
a particular form for the function in \eqref{nocutoff}, even if one wants to allow for the generality of the cutoff;  
This leads to the various HDE theories being motivated by different physical and mathematical considerations. In our work here,  we would like to consider the Hubble horizon cutoff, and our reasoning for making this choice is two-fold. On the one hand, we are now interested in the simplest way fractional HDE (FHDE, as we can refer to it later on) can produce universal evolution. In that regard, the Hubble Horizon cutoff provides a suitable choice. On the other hand, 
the Hubble horizon cutoff has been shown 
not to work within various frameworks \cite{Myung:2007pn,Li:2004rb}
and so if FHDE can provide an observationally consistent scenario 
then, it would represent a step forward in the ongoing appraisal of the Hubble cutoff.
\\
\\
With this in mind,  the definition \eqref{fracentropy} with the Hubble Horizon cutoff $L \to H^{-1}$ gives \begin{equation} \label{h1frarho}
 \rho_{DE} = 3 c^2 H^{\frac{3 \alpha - 2}{\alpha}}  . 
\end{equation} We also define the fractional density parameters for DE and DM as follows, using \eqref{fracentropy}: \begin{equation} \label{omegas}
    \Omega_{DE} = \frac{\rho_{DE}}{3 H^2} \to  c^2 H^{\frac{\alpha - 2 }{\alpha}} ,\quad \Omega_{m} = \frac{\rho_{m}}{3 H^2}.
\end{equation}
We also note, using \eqref{fried}, that \begin{equation}
    \Omega_{DE} + \Omega_{m} = \Omega_{DE} ( 1 + y)  = 1 , 
\end{equation}
where $y = \frac{\Omega_{DE}}{\Omega_{m}}$ and using this, alongside \eqref{fried}, \eqref{contm}, we can write \begin{equation} \label{h11}
    \frac{\dot{H}}{H^2} = - \frac{3}{2} ( 1 + w_{DE} + y) \Omega_{DE}.
\end{equation}
We can also utilize \eqref{contd} and \eqref{h1frarho} to obtain \begin{equation} \label{h12}
    \frac{\dot{H}}{H^2} = -\frac{3 \alpha}{3 \alpha - 2} (1 + w_{DE}).
\end{equation}
Equating \eqref{h11} and \eqref{h12} 
make it possible to write that \begin{equation} \label{weq}
    w_{DE} = -1 + \frac{(3 \alpha -2) (1 -\Omega_{DE})}{2 \alpha - \Omega_{DE} (3 \alpha - 2)},
\end{equation}
where we employed
the fact that $y = \frac{1 - \Omega_{DE}}{\Omega_{DE}} $. We can then use this parametrization for the dark 
EoS in \eqref{h12}. We thus  
find \begin{equation} \label{h13}
    \frac{\dot{H}}{H^2} = - \frac{3 \alpha (1-\Omega_{DE})}{2 \alpha - \Omega_{DE} (3 \alpha - 2)}.
\end{equation}
From this, one can directly get the deceleration parameter as \begin{equation} \label{qeq}
    q = -1 - \frac{\dot{H}}{H^2} = -1 + \frac{3 \alpha ( 1 - \Omega_{DE})}{2 \alpha - \Omega_{DE} (3 \alpha - 2)}.
\end{equation} Beyond this, one can then define the derivative of $\Omega_{DE}$ using \eqref{omegas} as \begin{equation}
    \Omega_{DE}' = \frac{\alpha -2 }{\alpha} \left( \frac{\Omega_{DE} \dot{H}}{H^2} \right),
\end{equation}
with overprime denoting differentiation with respect to $\ln{a} $. Using \eqref{h13} it is then easy to write that \begin{equation} \label{diffom}
    \Omega_{DE}' = \frac{3 (\alpha -2) (1-\Omega_{DE}) \Omega_{DE}}{2 \alpha  - \Omega_{DE}(3 \alpha - 2)}.
\end{equation}
We will now use the relation, $d \ln{a} = - \frac{dz}{1 + z}$ to find \begin{equation}
    (1+z) \frac{d \Omega_{DE}}{dz} = \frac{3 (\alpha -2 ) (1 - \Omega_{DE}) \Omega_{DE}}{2 \alpha - \Omega_{DE} (3 \alpha - 2) }.
\end{equation}
We can solve this differential equation to arrive at \begin{equation} \label{omegeq}
    \left( \frac{\Omega_{DE}}{\Omega_{DE0}} \right)^{2 \alpha} \left( \frac{1 - \Omega_{DE0}}{1 - \Omega_{DE}} \right)^{2 - \alpha} = (1+z)^{3(\alpha -2)},
\end{equation}
where $\Omega_{DE0}$ refers to $\Omega_{DE}$ at z=0. We shall now consider $\Omega_{DE0} \sim 0.69$, which is in line with recent observations of the universe \cite{adame2024desi3,adame2024desi4,adame2024desi6} and using this, we shall numerically solve \eqref{omegeq}. 
\begin{figure}[H]
    \centering
    \fbox{\includegraphics[width=.8\textwidth,height=.4\textwidth]{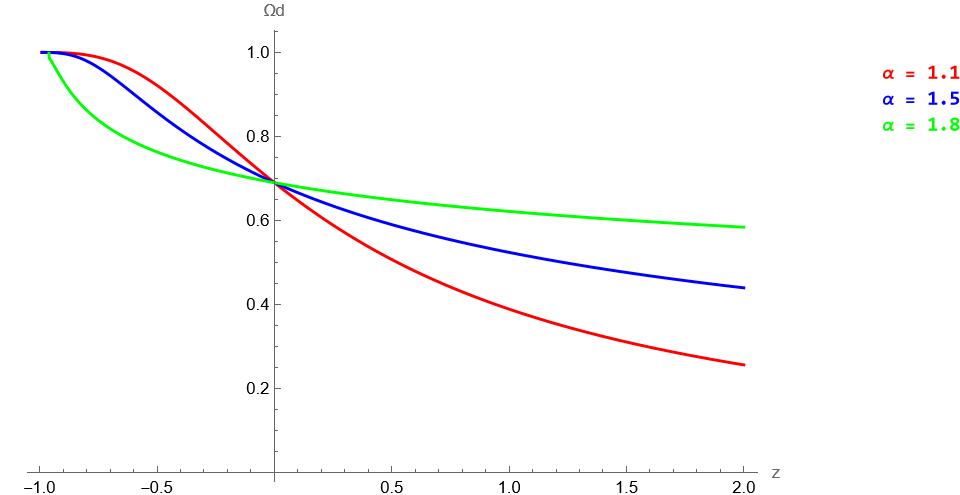}}
    \caption{Evolution of $\Omega_{DE}$ against the redshift z for various values of $\alpha$}
    \label{omegadfig}
\end{figure}

In figure \ref{omegadfig}, we have plotted the density parameter $\Omega_{DE}$ against the redshift z for various values of $\alpha$. One can retrieve 
that a suitable evolution of the DE energy density is 
possible to identify 
as 
effects from fractional features 
start to dominate.
More concretely, it is so because, for larger values of $\alpha$, like $\alpha= 1.8$ in Figure 1, the evolution of DE isn't very dynamic. By a dynamic nature here, we refer to the DE density increasing appreciably as the redshift becomes smaller. So, for larger values of $\alpha$, one sees that the growth of DE (and thereby the dark sector) is quite stagnant, while for smaller values, it is not the case. Indeed,  as one proceeds to 
smaller and smaller values of $\alpha$, like 1.5 and 1.1, we start to see that the cosmological evolution becomes more and more dynamic, dominating gradually as the redshift decreases,  which is in line with our expectation for FHDE, particularly when the fractional effects are most strong with e.g., $\alpha = 1.1$.  
\begin{figure}[H]
    \centering
    \fbox{\includegraphics[width=.8\textwidth,height=.4\textwidth]{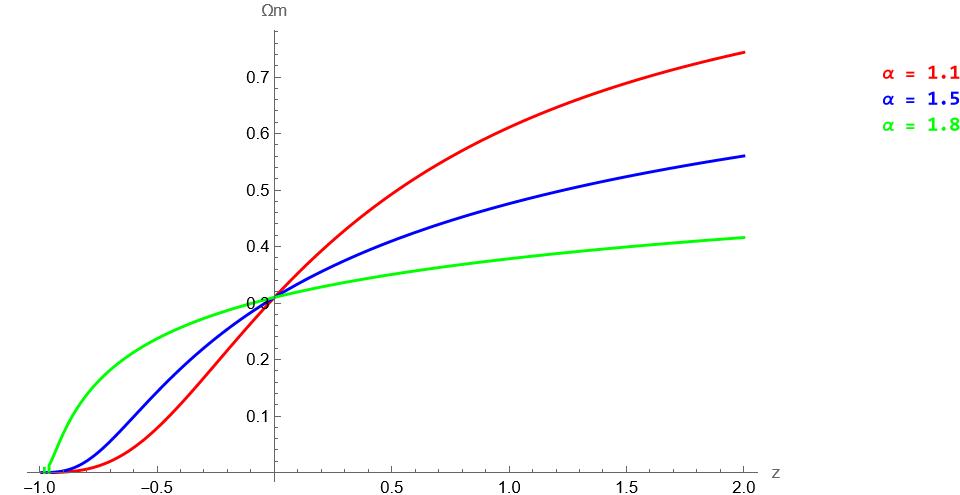}}
    \caption{Evolution of $\Omega_{m}$ against the redshift z for various values of $\alpha$}
    \label{omegamfig}
\end{figure}
We similarly plotted for the dark matter density parameter $\Omega_{m}$ in figure \ref{omegamfig}. We have again plotted here for various values of $\alpha$ as we did in the first figure, and once again sees that the evolution of DM is quite dynamic for smaller values of $\alpha$. In particular, one would note that figures \ref{omegadfig} and \ref{omegamfig} 
inform about
DE domination in the far future as one goes towards the limit $z \to -1$. 
\begin{figure}[H]
    \centering
    \fbox{\includegraphics[width=.8\textwidth,height=.4\textwidth]{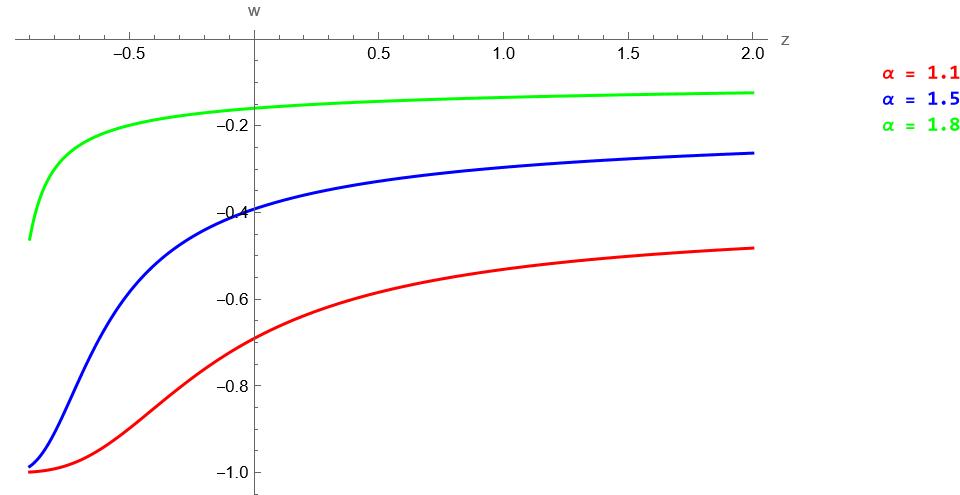}}
    \caption{Evolution of dark energy EoS parameter against the redshift z for various values of $\alpha$}
    \label{wfig}   
\end{figure}
The dark energy EoS parameter \eqref{weq} represents a significant quantity in studying the DE evolution, and we have plotted for that in figure \ref{wfig}. We took
various values of $\alpha$, and we see that the EoS parameter starts to show a behaviour more suitable and in line with observations for
small values of $\alpha$. In particular, 
for $\alpha=1.1$ the EoS parameter takes a value at $z=0$ which is in close agreement with the recent constraints on $w$ by DESI, see for example \cite{adame2024desi3,adame2024desi4,adame2024desi6}. As one goes towards larger values of $\alpha$, the evolution of the EoS parameter starts to deviate more and more from what one would expect for DE. We note here that for both $\alpha = 1.1$ and $\alpha=1.5$, the EoS parameter remains in the quintessence regime across the redshift range $2\leq z \leq -1$ and takes the value as $-1$ in the far future limit $z \to -1$. 
\begin{figure}[H]
    \centering
    \fbox{\includegraphics[width=.8\textwidth,height=.4\textwidth]{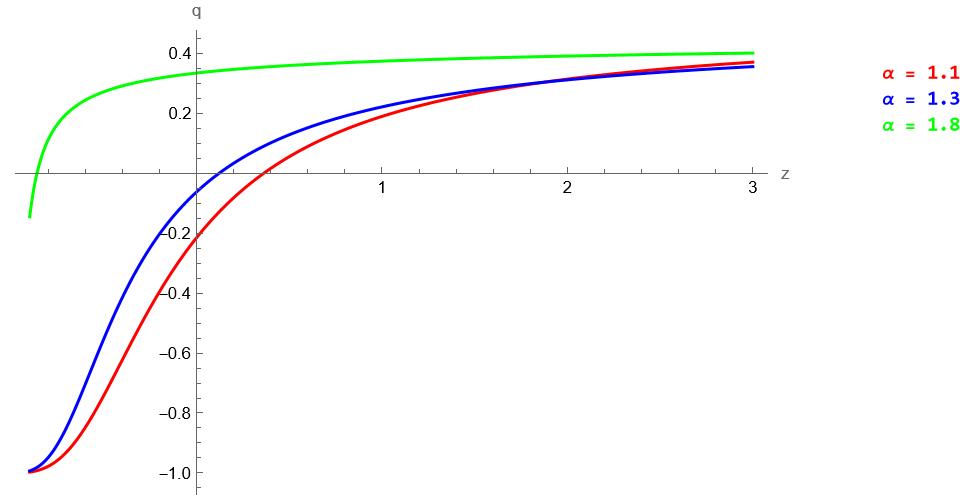}}
    \caption{Evolution of deceleration parameter q against the redshift $z$ for various values of $\alpha$}
    \label{qfig}      
\end{figure}
We have plotted similarly for the deceleration parameter $q$ \eqref{qeq} in figure \ref{qfig}. We note here that in the case of smaller values of $\alpha$, one sees an appropriate transition of the universe from the deceleration to the acceleration phase. In particular, for the case $\alpha=1.1$ one sees the transition redshift $z_{T} \sim 0.38$ which is well within the most recent constraints on $z_{T}$ \cite{Kumar:2022mtx}. The case of $\alpha=1.3$ presents $z_{T} \sim 0.17$, which would be pretty hard to be realistic. We note that for larger values of $\alpha$, with $\alpha = 1.8$ here, one sees that the transition 
to the acceleration phase of the universe seems to be pointed in happening in the future $z<0$, which is, of course, unrealistic. This again shows that smaller values of $\alpha$, where there is a stronger influence of 
fractional effects present a viable HDE scenario and are consistent with a Hubble horizon cutoff, which we find most satisfactory. 
\begin{figure}[H]
    \centering
    \fbox{\includegraphics[width=.8\textwidth,height=.4\textwidth]{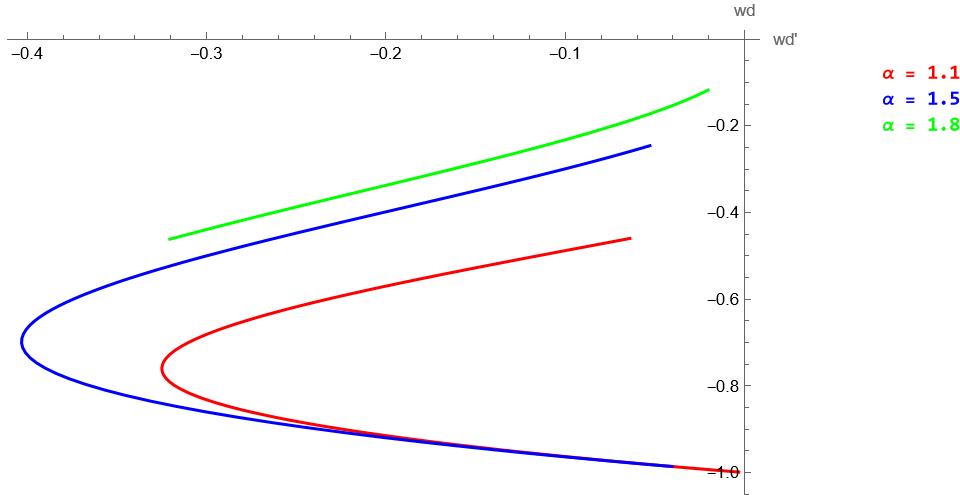}}
    \caption{The $w_{d}-w_{d}' $ plane for the redshift range $-1\leq z \leq 2$, for various values of $\alpha$}
    \label{wdwd'}      
\end{figure}

A beneficial way to understand the nature of models of dark energy which show a quintessence nature is by studying the nature of the $w_{d}-w_{d}'$ plane of that model, which was introduced in \cite{Caldwell:2005tm}. We note here that $w_{d}$ refers to the EoS parameter for DE, while $w_{d}'$ refers to its derivative with respect to $\ln{a}$. The plane allows one to distinguish between a ``freezing" or a ``thawing" form of quintessence regime for the model at hand. For regions in the plane where $w_{d} < 0$ and $w_{d}' > 0$, the model is in the thawing regime, while for $w_{d} < 0$ and $w_{d}' < 0$ the model is in the freezing regime. As one sees in figure \ref{wdwd'}, the FHDE scenario has a freezing form of quintessence throughout the range of $\alpha$. However, the implication is only particularly applicable for $\alpha=1.1$ and $1.5$ as these are regions for which the EoS remains in the quintessence regime, as seen earlier in figure \ref{wfig}.  
\begin{figure}[H]
    \centering
    \fbox{\includegraphics[width=.8\textwidth,height=.4\textwidth]{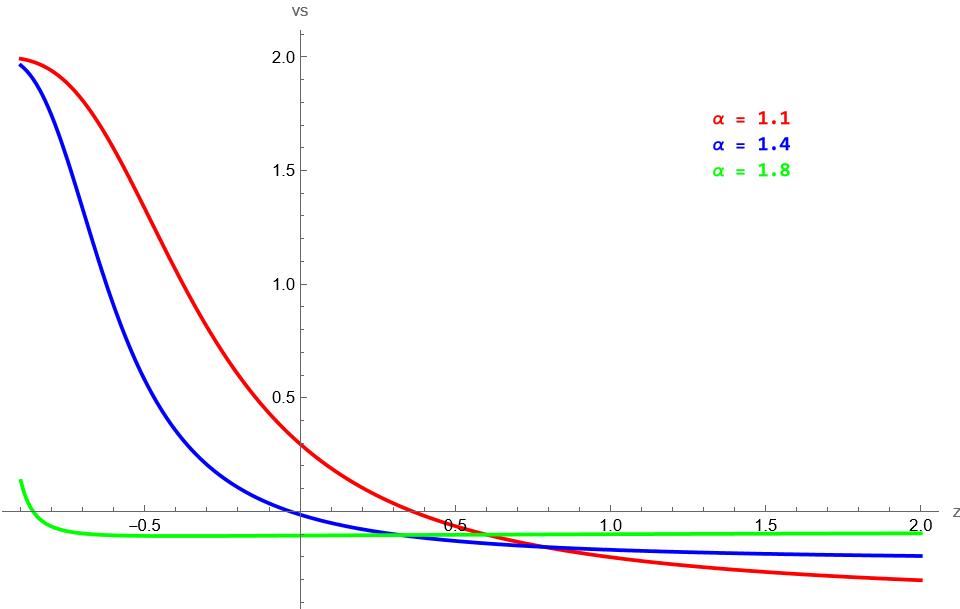}}
    \caption{The evolution of the squared sound speed $v_{s}^2$ against redshift z for various values of $\alpha$}
    \label{vsfig}      
\end{figure}

Finally, another important quantity with regards to HDE models, in particular, is the squared sound speed, given by 
\begin{equation}
    v_{s}^2 = \frac{\dot{p}}{\dot{\rho}} = w + \frac{\dot{w} \rho}{\dot{\rho}}.
\end{equation}
Using \eqref{h13} and the relation $d\ln{a} = -\frac{dz}{(1+z)}$, it is easy to write \begin{equation}
    v_{s}^2 = w + \frac{(1+z)(2 \alpha - \Omega_{DE} ( 3 \alpha -2)}{(3 \alpha - 2) ( 1 - \Omega_{DE}} \frac{dw}{dz}.
\end{equation}
One can then use \eqref{weq} and \eqref{omegeq} to arrive towards expressions for the sound speed in terms of $\Omega_{DE}(z)$ and using that, we have plotted the sound speed for various values $\alpha$ in figure \ref{vsfig}. It should be noted that classical instability has been a prevalent issue in HDE models \cite{Myung:2007pn}, a problem observed not only in conventional HDE but in various cutoffs and other extended HDE densities. In a lot of cases, it so happens that the model is always unstable towards classical perturbations, which is the case when the sound speed is not in the interval $0 \leq v_{s}^2 \leq 1$. It is important to note that the squared sound speed is usually between $0$ and $1$, as this region makes the most physical sense. While it may be straightforward to figure out why squared sound speeds cannot be negative, it is essential to clarify that sound speeds greater than 1 signify superluminal speeds of perturbations, thus pointing towards instability of the underlying theory against classical perturbations. While superluminal perturbations have been used as a possible way to explain issues in both the early and late-universe \cite{Bessada:2009ns,Afshordi:2006ad}, one usually sees this behaviour only in models with non-canonical Lagrangians and making the case for them even in those scenarios is quite difficult. We note in our model that our model maintains classical stability in the current epoch $z=0$ for smaller values of $\alpha$ where there is a more significant influence of the fractional effects. At the same time, in the far future, instability creeps in.  
We suspect this instability can be treated if one considers cutoffs besides the Hubble horizon cutoff for the model. Still, we shall not pursue that here and would leave that as an open direction for work for the interested reader. We have not specified any value for the constant $c$ in \eqref{fracrho}, so our results hold for any value of $c$.

\section{Conclusions}

In this paper, we have provided a new holographic dark energy scenario, which we have named Fractional Holographic Dark Energy. The name is so because this HDE scenario is motivated by 
corrections due to fractional calculus and fractional quantum mechanics, thus incorporating ideas from an exotic interface of mathematics and physics into gravitational (and cosmological) settings. The key feature of our model is a new energy density form for holographic dark energy, which we showed reduces to the conventional HDE energy density in the limit $\alpha \to 2$; NB. The parameter $\alpha$ is associated with fractional calculus features emerging in fractional (classical and quantum) mechanics and its implications.\\
\\
In particular, we derived this energy density from the holographic inequality, incorporating effects from the corrections to the Bekenstein-Hawking entropy with the assistance of the fractional Wheeler-De Witt equation. The energy density smoothly extends the conventional HDE energy density and provides outlooks to explore. In this line, we studied the cosmological evolution using the Hubble horizon cutoff for HDE. Then, we plotted the density parameters for DE and DM, the deceleration parameter, and the DE EoS parameter to give a detailed account of how the universe evolves in this scenario. The results are exciting, as our ``FHDE" model provides a very appropriate description of late-time cosmology. In particular, the Hubble horizon cutoff has been long shown to have quite a few issues when employed to study HDE. Still, our model can provide an acceptable 
evolution with this cutoff. Concretely, we showed that for smaller values of $\alpha$, one sees acceptable evolution. We also explored more subtleties of the model by studying its stability towards classical perturbations using the squared sound speed and detailing the nature of quintessence followed in it by considering the $w_{d}-w_{d}'$ plane evolution through a sizeable redshift range. Let us reiterate that we worked in the Hubble horizon cutoff, thereby showing that the influence of fractional effects can thus provide a  fairly 
consistent HDE scenario and, in particular, with a Hubble horizon cutoff. 
\\
\\
We can propose  multiple new directions of study based on our work here. For example, it would be interesting to see how this model fits  with the event horizon or the particle horizon cutoff instead of the Hubble horizon cutoff. One can also go towards more involved cutoffs like the Granda-Oliveros or the generalized Nojiri-Odintsov cutoff and study this scenario there. It will be interesting to study various thermodynamic aspects of this model, discussing the validity of the generalized second law with this model in the multiple cutoffs. One could also be tempted to study the far future scenarios of the universe, including rips, with this model (like in \cite{trivedi2024new}). Another point of interest can be studying various cosmological singularities \cite{trivedi2024recent,deHaro:2023lbq} in this scenario or studying these HDE models in non-GR cosmologies. These are all exciting endeavours, but we shall keep these directions open for future exploration as they currently lie beyond the scope of this work.   
\\
\\
\section*{Acknowledgements}
The authors would like to thank Robert Scherrer and Sergei Odintsov for various useful discussions on HDEs. PM acknowledges the FCT grant UID-B-MAT/00212/2020 
at CMA-UBI
plus the COST Action CA23130 (Bridging high and low energies in search of quantum gravity (BridgeQG)). 

\bibliography{references}
\bibliographystyle{unsrt}

\end{document}